\documentclass[prx,prl,a4paper,aps,twocolumn,superscriptaddress,longbibliography]{revtex4-2}
\usepackage{amssymb,amsmath,amsfonts,bm, bbm}
\usepackage{graphicx}
\usepackage{epstopdf}
\usepackage{times}
\usepackage{float}
\usepackage{lipsum}
\usepackage{color}
\usepackage{dcolumn}
\usepackage{bm}
\usepackage{subfigure}
\usepackage[colorlinks,linkcolor=blue,anchorcolor=blue,urlcolor=blue,citecolor=blue]{hyperref}
\usepackage{verbatim} 
\usepackage{amsmath} 
\usepackage{tikz}
\usetikzlibrary{quantikz}
\usepackage{braket}
\usepackage{bbold}
\usepackage{xypic}
\usepackage{amsthm}
\usepackage[most]{tcolorbox}

\theoremstyle{plain}
\newtheorem{thm}{\protect\theoremname}
\newtheorem{prop}[thm]{Proposition}

\newtheorem{cor}{Corollary}

\providecommand{\theoremname}{Theorem}
\newcommand*{\myproofname}{Proof}

\newcommand*{\id}{\mathbbm{1}}
\newcommand*{\Tr}{\textrm{Tr}}

\def\kk{\rangle\!\rangle}

\def\map{\mathcal}

\usepackage{hyperref}
\hypersetup{colorlinks=true, linkcolor=blue, citecolor=red, urlcolor=blue  }

\usepackage{physics}

\begin{document}

\title{Quantum Causal Inference with Extremely Light Touch}

\author{Xiangjing Liu }
\email{liuxj@mail.bnu.edu.cn}
\affiliation{CNRS@CREATE, 1 Create Way, 08-01 Create Tower, Singapore 138602, Singapore }
\affiliation{MajuLab, CNRS-UCA-SU-NUS-NTU International Joint Research Unit, Singapore}
\affiliation{Department of Physics, City University of Hong Kong, 83 Tat Chee Avenue, Kowloon, Hong Kong}
\affiliation{Department of Physics, Southern University of Science and Technology, Shenzhen 518055, China}
\affiliation{Centre for Quantum Technologies, National University of Singapore, Singapore 117543, Singapore}

\author{Yixian Qiu}
\affiliation{Centre for Quantum Technologies, National University of Singapore, Singapore 117543, Singapore}

\author{Oscar Dahlsten}
\email{oscar.dahlsten@cityu.edu.hk}
\affiliation{Department of Physics, City University of Hong Kong, 83 Tat Chee Avenue, Kowloon, Hong Kong}
\affiliation{Shenzhen Institute for Quantum Science and Engineering, Southern University of Science and Technology, Shenzhen 518055, China}
\affiliation{Institute of Nanoscience and Applications, Southern University of Science and Technology, Shenzhen 518055, China}

\author{Vlatko Vedral}
\affiliation{Clarendon Laboratory, University of Oxford, Parks Road, Oxford OX1 3PU, United Kingdom}


\begin{abstract}  
We give a causal inference scheme using quantum observations alone for a case with both temporal and spatial correlations: a bipartite quantum system with measurements at two times. The protocol determines compatibility with 5 causal structures distinguished by the direction of causal influence and whether there are initial correlations. We derive and exploit a closed-form expression for the space-time pseudo-density matrix (PDM) for many times and qubits. This PDM can be determined by light-touch coarse-grained measurements alone. We prove that if there is no signalling between two subsystems, the reduced state of the PDM cannot have negativity, regardless of initial spatial correlations. In addition, the protocol exploits the time asymmetry of the PDM to determine the temporal order. The protocol succeeds for a state with coherence undergoing a fully decohering channel. Thus coherence in the channel is not necessary for the quantum advantage of causal inference from observations alone.
\end{abstract}

\maketitle

\noindent {\bf{\em \large{Introduction}}}\\
Identifying cause-effect relations from observed correlations is at the core of a wide variety of empirical science~\cite{reichenbach1956direction, pearl2009causality}. Determining the causal structure, i.e.\  which variables influence others, is known as causal inference.  Causal inference is well-known to be important in understanding medical trials~\cite{balke1997bounds, prosperi2020causal}, and also appears in a range of machine learning applications~\cite{peters2017elements}. For example, by understanding the causal factors that give rise to different linguistic patterns, machine learning models can be trained to generate more accurate and meaningful text~\cite{feder2022causal}.

Causal inference can in principle be undertaken via {\em intervening} in the system~\cite{pearl2009causality,balke1997bounds}. Intervening to set a random variable to particular values in a controlled manner can be used to determine what other random variables that random variable influences. At the same time, e.g.\ in medical contexts~\cite{prosperi2020causal}, interventions may be costly or infeasible, motivating investigations into partial causal inference from {\em observations}~\cite{pearl2009causality,angrist1996identification,greenland2000introduction}. 

Similar questions have recently emerged concerning causal relations in quantum processes~\cite{leggett1985quantum, oreshkov2012quantum,brukner2014quantum,fitzsimons2015quantum,barrett2021cyclic,barrett2019quantum,hardy2005probability,chiribella2009theoretical,Milz2022resourcetheoryof,costa2016quantum,allen2017quantum,PhysRevA.79.052110,liu2022thermodynamics,parzygnat2022time,wolfe2021quantum,wolfe2020quantifying}. Interventions, like resetting the state of quantum systems, have been considered~\cite{ried2015quantum,bai2022quantum,chiribella2019quantum,maclean2017quantum,chaves2018quantum,agresti2022experimental, gachechiladze2020quantifying,nery2018quantum,agresti2020experimental}. It is known that in the classical case, observations alone are in general not sufficient to perform causal inference, which is connected to the famous phrase `correlation does not imply causation'. A natural question is therefore to identify minimal interventions and observations needed to determine causal relations in the quantum case~\cite{ried2015quantum}. It remains an open question to what extent observations (measurements) in the quantum case, which come together with an inescapable small disturbance, are sufficient for causal inference.   How `light-touch' can quantum causal inference be?

\begin{figure}
  \centering
  \includegraphics[scale=0.35]{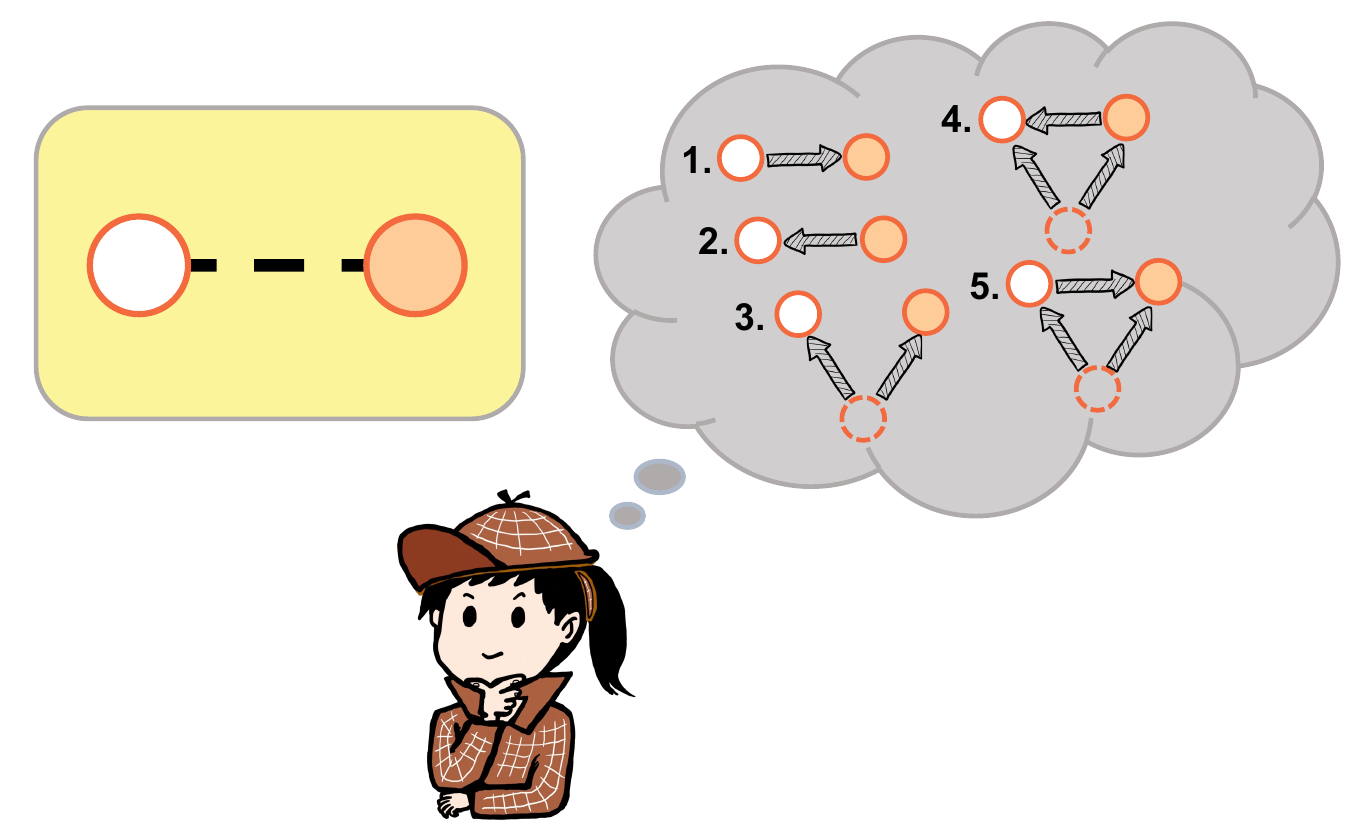}
  \caption{{\bf Quantum causal inference problem.} The observer gains data from observing two quantum systems $A$ (white ball) and $B$ (red ball) which is correlated. In line with Reichenbach's principle, we allow for five possible causal structures: 1) $A$ has direct influence on $B$; 2) $B$ has a direct influence on $A$; 3) there is a common cause (dashed ball) acting on $A$ and $B$, meaning correlations in the initial state; 4) a combination of cases 1 and 3; 5) a combination of cases 2 and 3. The observer wants to determine which of those possible causal structures is the case. 
 }
  \label{fig:qsetup}
\end{figure}

We here address this question for the case of bipartite quantum systems of arbitrary numbers of qubits and measurements at two times. To be precise, we formulate the quantum causal inference problem as follows.  As shown in FIG.~\ref{fig:qsetup}, the observer has data from observing two quantum systems $A$ and $B$. The observer wants to know the causal structure of the process of generating the data. In line with Reichenbach's principle~\cite{reichenbach1956direction} we allow for five causal structures that are to be distinguished (see FIG.~\ref{fig:qsetup}). These structures are distinguished by the direction of any causal influence between $A$ and $B$, and by whether there are initial correlations or not. Scenarios with causal influence in both directions (loops), such as global unitaries on $A$ and $B$ are excluded, such that there is a well-defined causal direction~\cite{reichenbach1956direction,pearl2009causality} (nevertheless many of our results apply to such cases).  We devise an explicit scheme for determining which causal structures are compatible with the data. The scheme is derived via the pseudo-density matrix (PDM) formalism, which assigns a PDM to the data table of experiments involving measurements on systems at several times~\cite{fitzsimons2015quantum}. Firstly one identifies whether there is negativity in certain reduced states of the PDM. Then one evaluates the time-asymmetry of the PDM. The scheme employs no reset-type interventions but rather only coarse-grained projective measurements, thereby proving that causal inference can indeed be achieved with very light touch in the quantum case. 

We proceed as follows. After introducing the PDM formalism we present the main theorems, the protocol and an example. Further details are provided in the Supplementary Information.

\vspace{0.5cm}
\noindent {\bf {\em \large{Results}}}\\
\noindent {\bf {\em PDM formalism for measurements at multiple times, systems.}}
The pseudo-density matrix (PDM) formalism, developed to treat space and time equally~\cite{fitzsimons2015quantum}, provides a general framework for dealing with spatial and causal (temporal) correlations. Research on single-qubit PDMs has yielded fruitful results~\cite{horsman2017can,fullwood2022quantum, jia2023quantum,marletto2021temporal, zhao2018geometry, marletto2019theoretical,zhang2020different,zhang2020quantum,pisarczyk2019causal}. For example, recent studies have utilized quantum causal correlations to set limits on quantum communication~\cite{pisarczyk2019causal} and to understand how dynamics emerge from temporal entanglement~\cite{marletto2021temporal}. Furthermore, the PDM approach has been used to resolve causality paradoxes associated with closed time-like curves~\cite{marletto2019theoretical}.

The PDM generalises the standard quantum $n$-qubit density matrix to the case of multiple times. The PDM is defined as 
\begin{align}\label{eq: defPDM}
R_{1...m}=\frac{1}{2^{mn}} \sum^{4^n-1}_{i_1=0}...\sum^{4^n-1}_{i_m=0}  \langle  \{ \tilde{\sigma}_{i_{\alpha}} \}^m_{\alpha=1} \rangle   \bigotimes^m_{\alpha=1} \tilde{\sigma}_{i_{\alpha}},
\end{align}
where $  \tilde{\sigma}_{i_\alpha}\in \{ \sigma_0,\sigma_1,\sigma_2, \sigma_3\}^{\otimes n} $ is an $n$-qubit Pauli matrix at time $t_\alpha$. $\tilde{\sigma}_{i_\alpha}$ is extended to an observable associated with up to $m$ times, $\bigotimes^m_{\alpha=1} \tilde{\sigma}_{i_{\alpha}}$ that has expectation value  $\langle  \{ \tilde{\sigma}_{i_\alpha} \}^m_{\alpha=1} \rangle $. We shall return later to what measurement this expectation value corresponds to. The standard quantum density matrix is recovered if the Hilbert spaces for all but one time, say $t_{\alpha'}$ are traced out, i.e.,\ $\rho_{\alpha'}=\Tr_{\alpha \neq \alpha'}R_{1...m}$. The PDM is Hermitian with unit trace but may have negative eigenvalues.

The negative eigenvalues of the PDM appear in a measure of temporal entanglement known as a causal monotone $f(R)$~\cite{fitzsimons2015quantum}. Analogously to the case of entanglement monotones~\cite{vidal2000entanglement}, in general $f(R)$ is required to satisfy the following criteria: (I) $f(R) \geq 0$, (II) $f(R)$ is invariant under local change of basis, (III)  $f(R)$ is non-increasing under local operations, and (IV) $ \sum_i p_i f(R_i) \geq f(\sum p_i R_i)$.
Those criteria are satisfied by~\cite{fitzsimons2015quantum}
\begin{align}\label{eq: causmono}
f(R) :=||R||_{tr}-1 = \Tr \sqrt{ R R^\dag}-1.
\end{align} If $R$ has negativity, $f(R)>0$.  An intuition for why $f(R)$ serves as a sign of causal influence is that negative eigenvalues tell you that the PDM is associated with measurements at multiple times; in the case of a single time, there would be a standard density matrix with no negativity.

The PDM negativity $f(R)$ can thus be used to distinguish, at least in some cases, whether the PDM corresponds to two qubits at one time or one qubit at two times. This can be viewed as a simple form of causal inference, raising the question of whether the inference involving two parties (of multiple qubits) at multiple times depicted in FIG.~\ref{fig:qsetup} can be undertaken in a similar manner. A key challenge in this direction is to find a closed-form expression for the PDM $R$, from which one can see whether $f(R)>0$. 

\noindent {\bf {\em Closed form for m-time n-qubit PDMs.}} We derive a closed-form expression for the PDM for $n$ qubits and $2$ times, before generalising the expression to $m$ times. 

Consider the PDM of $n$ qubits undergoing a channel $\mathcal{M}_{2|1}$ between times $t_1$ and $t_2$. In order to fully define the PDM of Eq.~\eqref{eq: defPDM} it is necessary to further define how the Pauli expectation values  $\langle  \{ \tilde{\sigma}_{i_{\alpha}} \}^m_{\alpha=1} \rangle$ are measured, since that choice impacts the states in between the measurements. We, importantly, choose {\em coarse-grained} projectors
\begin{align}
\label{eq:coarseP}
\Bigl\{& P^{\alpha}_+= \frac{ \mathbbm{1}  + \tilde{\sigma}_{i_\alpha}}{2} , 
P^{\alpha}_-= \frac{\mathbbm{1}  - \tilde{ \sigma}_{i_\alpha}}{2} \Bigl\},
\end{align} 
where $\alpha$ in $i_\alpha$ labels the time of the measurement. These are coarse-grained in the sense of being sums of rank-1 projectors, and by inspection generate lower measurement disturbance than fine-grained projectors in general. The coarse-grained projectors' probabilities determine the expectation values $\langle  \{ \tilde{\sigma}_{i_{\alpha}} \}^m_{\alpha=1} \rangle$. (See the Supplementary Information for a circuit to implement these measurements.)

The closed form of the PDM that we shall derive employs the Choi-Jamio{\l}kowski (CJ) matrix of the completely positive and trace-preserving (CPTP) map $\mathcal{M}_{2|1}$,~\cite{choi1975completely,jamiolkowski1972linear}. An equivalent variant of the definition of the CJ matrix is as follows:
\begin{equation} \label{eq: CHOI}
M_{12}:= \sum^{2^n-1}_{i ,j=0}\ket{i}\bra{j}^T \otimes  \mathcal{M}_{2|1} \left( \ket{i}\bra{j}  \right),
\end{equation}
where the superscript $T$ denotes the transpose.
We show (see the Supplementary Information) that the 2-time $n$-qubit PDM, under coarse-grained measurements, can be written in a surprisingly neat form in terms of $M_{12}$.
\begin{thm}
\label{thm: 2timesPDM}
Consider a system consisting of $n$ qubits with the initial state $\rho_1$.  The coarse-grained measurements of Eq.~\eqref{eq:coarseP} are applied at times $t_1$ and $t_2$.  The channel $\mathcal{M}_{2|1}$ with CJ matrix $M_{12}$ is applied in-between the measurements. The $n$-qubit PDM can then be written as
\begin{align} \label{eq:2timenqubit}
R_{12} = \frac{1}{2}( M_{12} \, \rho  +\rho  \, M_{12}  ) ,
\end{align}
where $\rho :=\rho_1 \otimes \mathbbm{1}_2$.
\end{thm}
Theorem  \ref{thm: 2timesPDM} extends an earlier known form for the single qubit case to multiple qubits that may have entanglement ~\cite{horsman2017can, zhao2018geometry}. The theorem provides an operational meaning for a mathematically motivated spatiotemporal formalism~\cite{parzygnat2022time}. Moreover, the $n$-qubit PDM will enable us to investigate phenomena that cannot be explored in the single qubit case, such as quantum channels with associated extra qubits constituting a memory~\cite{pisarczyk2019causal}. 

We next, for completeness, stretch the argument to multiple times. Consider initially an $n$-qubit state $\rho_1$ measured at time $t_1$, undergoing the channel $\mathcal{M}_{2|1}$, measured at time $ t_2$, undergoing $\mathcal{M}_{3|2}$ and measured at time $t_3$. The central objects to determine are the joint expectation values of the observables at three times. These can be written as 
\begin{align}
\label{eq:3timeexpectation}
 & \langle  \tilde{\sigma}_{i_1} , \tilde{ \sigma}_{i_2} , \tilde{\sigma}_{i_3} \rangle \nonumber\\
 =&\Tr_{23}  [ M_{23} \Bigl(  P^{2}_{ +} \rho_2^{(\tilde{\sigma}_{i_1})}  P^{2}_{ +} -  P^{2}_{ -} \rho_2^{(\tilde{\sigma}_{i_1})}  P^{2}_{ -}  \Bigl) \otimes \tilde{\sigma}_{i_3} ] ,
\end{align}
where we denote the CJ matrices for channels $\mathcal{M}_{2|1}, \mathcal{M}_{3|2}$ by $M_{12},M_{23} $ respectively, and (see the Supplementary Information)
\begin{align}
\label{eq:rho2}
\rho_2^{(\tilde{\sigma}_{i_1})}  =\Tr_1[  R_{12} \, \tilde{\sigma}_{i_1} \otimes \mathbbm{1}_2 ] .
\end{align}

Eqs.~\eqref{eq:3timeexpectation} and \eqref{eq:rho2} then together imply that 
\begin{align}
\label{eq:3timeexpectationR}
  &\langle  \tilde{\sigma}_{i_1} ,\tilde{ \sigma}_{i_2} , \tilde{\sigma}_{i_3} \rangle \nonumber\\
 = &\frac{1}{2} \Tr  [ ( M_{23}  R_{12}   +    R_{12}  M_{23}  )  \tilde{\sigma}_{i_1} \otimes \tilde{\sigma}_{i_2}  \otimes \tilde{\sigma}_{i_3} ] ,
\end{align} 
 where implicit identity matrices are now omitted for notational convenience.

From Eq.\eqref{eq:3timeexpectationR}, demanding that
\begin{align}
\label{eq:R123}
R_{123}=\frac{1}{2}  ( R_{12}   M_{23} +   M_{23} R_{12} ),
\end{align}
gives expectation values consistent with the PDM definition of Eq.~\eqref{eq: defPDM}. Since the expectation values 
{\em uniquely} determine the PDM, Eq.~\eqref{eq:R123} must be the correct expression. 

The above derivation can be directly generalized to more than three times:
\begin{thm}
The $n$-qubit PDM across $m$ times is given by the following iterative expression
\begin{align}
R_{12...m}= \frac{1}{2} (R_{12...m-1} M_{m-1,m}+  M_{m-1,m} R_{12...m-1}  )
\end{align}
with the initial condition $R_{12}= \frac{1}{2}( \rho \, M_{12} +M_{12} \, \rho ) $ where $M_{m-1,m}$ denotes the CJ matrix of the $(m-1)$-th channel.
\end{thm}
This iterative expression, proven in the Supplementary Information, can be written in a (possibly long) closed form sum in a natural manner. We have thus extended a key tool in the PDM formalism from the cases of single qubits, two times or two qubits single time to the case of $n$ qubits at $m$ times for any $n$ and $m$.

\noindent {\bf {\em Relation between PDM negativity and possibility of common cause.}} PDM negativity ($f>0$) was linked to cause-effect mechanisms for the case of one qubit at 2 times or 2 qubits at one time in Ref.~\cite{fitzsimons2015quantum}. We now consider the case of several qubits and several times, such that there may be combinations of temporal and spatial correlations.  We use Eq.~\eqref{eq:2timenqubit} to derive a relation between the negativity of parts of the PDM and the possibility of a common cause, meaning correlations in the initial state. 

\begin{figure}
  \centering
  \includegraphics[scale=0.8]{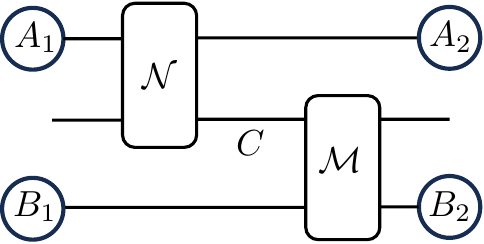}
  \caption{ {\bf{ Semicausal channel.}}   Semicausal channels are bipartite channels which can be decomposed into either $ \mathcal{M}_{BC} \circ \mathcal{ \mathcal{N}}_{AC} $ or $\mathcal{N}_{AC} \circ \mathcal{ \mathcal{M}}_{BC} $ where $C$ is an ancilla, as in the above circuit. The circles here indicate possible measurements. In this example, which is consistent (only) with cases 1, 3 and 5 in FIG.~\ref{fig:qsetup}, $A$ can causally influence $B$ while the inverse is not true.}   
  \label{fig:semicausal}
\end{figure}

We model the possible directional dynamics of Fig.~\ref{fig:qsetup} as so-called \textit{semicausal} channels~\cite{beckman2001causal, eggeling2002semicausal}. Semicausal channels are those bipartite completely positive trace-preserving (CPTP) maps that do not allow one party to signal or influence the other. If the channel $\map P$ does not allow $B$ to influence $A$,  it must admit the decomposition $\mathcal{P} = \mathcal{M}_{BC} \circ \mathcal{N}_{AC}$~\cite{eggeling2002semicausal}. The circuit representation of $ \mathcal{P}  $ on $A$ and $B$ across two times $t_1, t_2$, is depicted in FIG.~\ref{fig:semicausal}.  
The following theorem shows that when there is no signalling from $B$ at time 1 to $A$ at time 2, the PDM $R_{B_1A_2} $ has no negativity for any input state.
\begin{thm}[null PDM negativity for semicausal channels] \label{thm:nullneg}
If a quantum channel $\mathcal{P}$ does not allow signaling from $B$ to $A$, then, for any state $\rho_{A_1B_1}$ at time $t_1$,  the PDM $R_{B_1A_2} $ is positive semidefinite and the PDM negativity $f(R_{B_1A_2} )=0$.
\end{thm}
The theorem implies that {\em only} the existence of causal influence between $B$ and $A$ allows for  $f(R_{B_1A_2} )>0$. In particular, if there is no causation from $B_1$ to $A_2$, any initial correlations between $A_1$ and $B_1$ cannot make the PDM negativity $f(R_{B_1A_2})>0$. In contrast, several other observation-based measures such as the mutual information can be raised from initial correlations alone~\cite{janzing2013quantifying}. 

Theorem~\ref{thm:nullneg} additionally has value for the more restricted task of characterising whether channels are signalling, as considered in Refs.~\cite{beckman2001causal, eggeling2002semicausal}. If $f(R_{B_1A_2} )> 0$ the channel must be signalling from $B$ to $A$. In this restricted task one may vary over input states. There are reasons to believe pure product states may maximise  $f(R_{B_1A_2} )$ for a given channel.  From property IV of $f$, with a given $R=\sum_ip_iR_i$, the most negative pure state $R_{i*}:=\mathrm{argmax}\, f(R_i)$ respects $f(R_{i*})\geq f(R)$. We moreover conjecture that if the channel is signalling from B to A, we can always find a pure {\em product} input state such that $f(R_{B_1A_2} )> 0$. We prove this conjecture for a quite general case of 2-qubit unitary evolutions in the Supplementary Information.


\noindent {\bf {\em Exploiting time asymmetry to distinguish cause and effect.}} Consider the case where there is negativity $f(R_{AB}) >0 $ but it is not known which is the cause or effect, i.e.. the time-label is unknown.  We can then exploit the asymmetry of temporal quantum correlations~\cite{liu2023inferring} to distinguish the cause and effect, and to determine whether there is a common cause.   

The asymmetry of temporal quantum correlations can be defined by comparing forwards and time-reversed PDMs~\cite{liu2023inferring}. The time reversed PDM, 
\begin{equation}\label{eq:reversePDM}
\bar {R}_{AB}:= S \, R_{AB} \, S^\dag, 
\end{equation}
where $S$ denotes the $n$-qubit swap operator~\cite{liu2023inferring, parzygnat2022time}. The methods given here to find a closed-form expression for $R_{AB}$ can be similarly applied to show that  $\bar {R}_{AB} =\frac{1}{2}( \pi \, \bar{M} +\bar{M} \, \pi )$, where $\pi:= (\Tr_A R_{AB}) \otimes \id_A $ and $\bar M$ is the CJ matrix of the time reversed process. The CJ matrices $M$ and $\bar {M}$ can be extracted via a vectorization of $R$ and $\bar R$, respectively~\cite{liu2023inferring}.  Let $T$ denote the transpose on the initial quantum system. The Choi matrices of the process and its time reversal are given by $M^{T} $ and $\bar{M}^{T} $, respectively. A process being CP is equivalent to its Choi matrix being positive~\cite{choi1975completely,jamiolkowski1972linear}. When only one of the two Choi matrices is positive we say there is asymmetry of the temporal quantum correlations. 

The asymmetry can be used to distinguish different causal structures. If there is no initial correlation (no common cause) the forwards process is CP but in general the reverse process may be not positive semidefinite ($\bar{M}^T\ngeqslant 0$). Furthermore, if both Choi matrices  are not positive semidefinite  ($\bar{M}^T, M^T\ngeqslant 0$) then neither process is CP and there must be a common cause (initial correlations).

\noindent {\bf {\em Protocol for quantum causal inference.}} We will now make use of the results from previous sections to give a protocol that determines the compatibility of the experimental data with the causal structures shown in FIG.~\ref{fig:qsetup}.  In line with causal inference terminology~\cite{pearl2009causality}, we say that the data and a causal structure are  {\em compatible} if experimental data could have been generated by that structure. As in causal inference in general, compatibility is not guaranteed to be unique.

The causal structures of FIG.~\ref{fig:qsetup} are as follows. Case 1 is the cause-effect mechanism in one direction, when there are two instances of quantum systems $A$ and $B$ located in space and actions on $A$ influence the reduced state on $B$ and the actions on $B$ do not influence the reduced state on $A$. Case 2 is the same mechanism as Case 1 but in the opposite direction. Case 3 is the pure common cause mechanism, with no influence between $A$ and $B$. There is a common cause, meaning correlations at the initial time $t_1$, iff $R_{A_1B_1}\neq R_{A_1}\otimes R_{B_1}$.  Cases 4 and 5 is when there is a common cause mechanism and also a cause-effect mechanism. Cases 4 and 5 are distinguished by the directionality of the cause-effect mechanism.

Recall that the setting involves 2 systems $A$ and $B$ and 2 times $t_i$ and $t_j$. We are given the data that constructs the PDM $R_{A_iB_j}$ and assume that the data has correlations ($R_{A_iB_j}\neq R_{A_i}\otimes R_{B_j}$ for whatever $i$, $j$ we are given data for) so that there is a non-trivial causal structure. We are not given the data that constructs the PDM $R_{A_iB_iA_jB_j}$ and do not have enough data to reconstruct the full channel on $AB$ in general. We are moreover not told which time is measured first. The protocol is as follows: 
\begin{itemize}

\item[(1)]
 {\bf Evaluating compatibility with a common-cause mechanism}. Consider the case of no negativity ($f(R_{A_iB_j})=0$). Theorem~\ref{thm:nullneg} implies that only the existence of causal influence between $A_i$ and $B_j$ can allow for negativity. The purely common cause mechanism (case 3 in FIG.~\ref{fig:qsetup}, $R_{A_1B_1}\neq R_{A_1}\otimes R_{B_1}$) is in contrast compatible with no negativity. Thus for no negativity, the protocol is to conclude that the data $R_{A_iB_j}$ is compatible with the (purely) common cause mechanism. 

\item[(2)] 
 {\bf Evaluating compatibility with different cause-effect mechanisms.}  Consider the case of negativity ($f(R_{A_iB_j})>0$). Theorem~\ref{thm:nullneg} rules out the common cause mechanism and we are left to evaluate the compatibility of the data with cases 1, 2, 4, and 5 in FIG.~\ref{fig:qsetup}. We make use of the time asymmetry results described around Eq.~\eqref{eq:reversePDM} for this evaluation. In particular we extract the two Choi matrices $M^{T}$,  $\bar{M}^{T}$ associated with $R_{A_iB_j}$ and its time reversal $\bar{R}_{A_iB_j}$. The basic idea is that $M^{T}>0$ means there is a CP map on $A$ that gives $B$, indicating that $A$ could be the cause and $B$ the effect. More specifically, 
 \begin{itemize}
 \item If $M^{T} \geq 0$ and $\bar{M}^{T} \ngeqslant	0$, the data is compatible with $A \rightarrow B$ (case 1 in FIG.~\ref{fig:qsetup}).
 \item If $M^{T} \ngeqslant	 0$ and $\bar{M}^{T} \geq 0$, the data is compatible with $A \leftarrow B$ (case 2 in FIG.~\ref{fig:qsetup}).
 \item If $M^{T} \geq 0$ and $\bar{M}^{T} \geq 0$, the data is compatible with case 1 and/or case 2 in FIG.~\ref{fig:qsetup}.  
 \end{itemize}
 
\item[(3)] If none of the above conditions are satisfied, i.e.,  $f(R_{AB})> 0, M^{T} \ngeqslant	0$ and $\bar{M}^{T} \ngeqslant	 0$, the causal structure is compatible only with case 4 or 5 in FIG.~\ref{fig:qsetup}.
 
 \end{itemize}
Detailed justifications for the above protocol are given in the Supplementary Information. The Supplementary Information also contains a semi-definite program motivated by a technical subtlety when extracting the CJ matrix from the PDM. When both $\rho$ and $\pi$ are of full rank,  $M$ and $\bar M$ can be uniquely extracted using the vectorization technique. However, when they are rank deficient, there are infinitely many solutions for $M$ and $\bar M$. Ref.~\cite{song2024causal} also showed how solving for the process in the case where the marginal is rank deficient is a semidefinite problem for the case of a single qubit. Therefore, we design a semidefinite programming problem to find all possible CJ matrices where $M^{T_1} $ and $\bar{ M}^{T_1}$ are the least negative.

The protocol identifies compatibility and it is natural to wonder whether it uniquely identifies the structure used to generate the data. For at least part of the protocol this appears to be the case. Numerical simulations of 2-qubit cases show a near unit probability that if $f(R_{A_iB_j})>0$ the data is indeed not generated by the common cause mechanism (see Supplementary Information).

\noindent {\bf {\em Example: cause-effect mechanism.}}  We now consider an example that shows how our light-touch protocol can resolve the causal structure even for channels that do not preserve quantum coherence. Let systems $A$ and $B$ be uncorrelated single qubit systems and the end effect of the compound channel $ \mathcal{M}_{BC} \circ \mathcal{N}_{AC } $ on the compound system $AB$ be the channel that measures the system $A$, recording the outcome in $C$ and then preparing a state on system $B$ that depends on $C$, as in Fig.~\ref{fig:semicausal}. Denote the effective channel on $AB$ by $\mathcal{L}_{A \rightarrow B} =\Tr_{CA}\circ \mathcal{M}_{BC} \circ \mathcal{N}_{AC }$. For concreteness we choose $\mathcal{N}_{AC }(\rho_A\otimes \ket{0}_C\bra{0})=\bra{0} \rho_{A} \ket{0}  \ket{00}_{AC}  \bra{00} +\bra{1} \rho_{A} \ket{1}  \ket{11}_{AC} \bra{11}$  and $\mathcal{M}_{BC}(\rho_{B}\otimes \rho_C)=S(\rho_{B}\otimes \rho_C)S^{\dagger}$ where $S$ is the unitary swap. Thus the action of $\mathcal{L}_{A \rightarrow B}$ on the state is $\mathcal{L}_{A \rightarrow B} (\rho_{A}) = \bra{0} \rho_{A} \ket{0}  \ket{0}_{B}  \bra{0} +\bra{1} \rho_{A} \ket{1}  \ket{1}_{B} \bra{1}$. Therefore, the CJ matrix of $\mathcal{L}$ in the Pauli basis is 
 \begin{align}\label{eq: MPchoi}
L= \frac{1}{2} \sum^3_{i=0}   \sigma_i \otimes  \mathcal{L}( \sigma_i ) = \frac{1}{2} ( \sigma_0 \otimes  \sigma_0  + \sigma_3 \otimes \sigma_3). 
\end{align}
 Substituting Eq.~\eqref{eq: MPchoi} into Eq.~\eqref{eq:2timenqubit}, the PDM 
\begin{align}
R_{A_1B_2} 
 =&   \left(\frac{1}{2} \rho_{A_1}    +\frac{1}{4} \sigma_3  +  \frac{z}{4} \sigma_0  \right) \otimes \ket{0}\bra{0}  \nonumber\\
 &+  \left(\frac{1}{2} \rho_{A_1} - \frac{1}{4} \sigma_3 - \frac{z}{4} \sigma_0 \right) \otimes \ket{1}\bra{1}, 
\end{align}
where $z:= \Tr (\rho_{A_1} \sigma_3) $. The eigenvalues of $ \rho_{A_1}  +\frac{1}{2} \sigma_3 +  \frac{z}{2} \mathbbm{1}$ are $ \frac{1}{2}(1+z\pm \sqrt{(1+z)^2+x^2+y^2})$ with $x:=  \Tr( \rho_{A_1} \sigma_1), y:=  \Tr( \rho_{A_1} \sigma_2)$. When $x^2 +y^2=0$ , the PDM is positive ($f(R_{A_1B_2})=0$) without coherence in the Pauli-$z$ basis.  However, the PDM is negative ($f(R_{A_1B_2})>0$) exactly when $x^2 +y^2>0$, i.e., when the initial state $\rho_{A_1}$ is coherent in the Pauli-$z$ basis. 

For concreteness we now assume the initial state is given by $\rho_{A_1B_1}= \left[(1-\lambda) \frac{\id}{2} + \lambda \ket{+}\bra{+} \right] \otimes \ket{0}\bra{0}, \lambda \in (0,1) .$  The Choi matrix of the time reversal process (Eq.~\eqref{eq:reversePDM}) can be calculated to be 
\begin{align}
\bar{L}^{T} =
\frac{1}{2}\begin{pmatrix}
      2 & \lambda  \\  
      \lambda  &  0 
\end{pmatrix} \otimes \ket{0}\bra{0}+
\frac{1}{2}\begin{pmatrix}
      0 & \lambda  \\  
      \lambda  &  2 
\end{pmatrix} \otimes  \ket{1}\bra{1}.
\end{align}
Clearly, $L^{T}  \geq 0 $ and $ \bar{L}^{T} \ngeqslant 0 $ for any $\lambda\in (0,1)$.

Applying the causal inference to the above case we would firstly note $f(R_{A_1B_2})>0$ so case 3 is ruled out. Since $L^{T}  \geq 0 $ and $ \bar{L}^{T} \ngeqslant 0$, the data is compatible with $A\rightarrow B$ (case 1 in FIG.~\ref{fig:qsetup}).

The example has implications for when the apparent quantum advantage of not requiring interventions for causal inference exists. An earlier observational protocol~\cite{ried2015quantum} showed this advantage existing for a case of coherence preserving channels. The above example using our observational protocol indicates that coherence preserving channels is not required for this apparent quantum advantage. In the above example there is coherence in the initial state but a decoherent channel. A further example of applying the protocol to a cause-effect mechanism with a common cause is given in the Supplementary Information.

\vspace{0.5cm}
\noindent {\bf {\em \large{Discussion}}} \\
The results naturally point towards several developments: (i) Our closed form PDM may enable Leggett-Garg type inequalities, which concern 3 or more times~\cite{leggett1985quantum, VitaglianoB23}, to be extended to non-trivial evolutions; (ii) The causal inference protocol may be generalisable to networks of multiple times and parties using the closed form; (iii) The causality monotone might be possible to witness via observables, c.f.~\cite{araujo2015witnessing}; (iv) Other formalisms based around the CJ isomorphism could likely be employed analogously, and may offer alternative tools and perspectives~\cite{liu2023unification, oreshkov2012quantum,costa2016quantum, chiribella2009theoretical,allen2017quantum,ried2015quantum}; (v) Our scheme can be used to determine {\em classical} causal structures without interventions provided that these can be probed in quantum superposition, e.g.\ as in the case of typical optical table equipment; (vi) Why are such light-touch interventions sufficient for quantum causal inference? (vii) The protocol could be strengthened to distinguish between causal mechanisms 4 and 5 for the special case when there is negativity both in the PDM, the Choi matrix and the time-reversed Choi matrix; (viii) An important open question is whether the approach can be generalised to other measurement schemes.

 \vspace{0.5cm}
\noindent {\bf {\em \large{Methods}}} \\
\noindent{\bf { \em Coarse-grained measurement underlying closed-form PDM.}}
Let us take a two-qubit system to illustrate our design of measurement events.
At initial time $t_1$, we implement the observable $  \sigma_{i}^{A} \otimes \sigma_{j}^{B} $. This observable can be decomposed into linear combinations of projectors in several ways. For example, 
\begin{align}\label{eq: statistics}
 \sigma_{i} \otimes \sigma_{j}  &=   P_1   +   P_2   -    P_3   -    P_4  \nonumber\\
 &=   (P_1+P_2 ) -  (P_3+P_4) ,
\end{align}
where 
\begin{align}
&P_1 := \frac{1}{4} (\mathbbm{1} + \sigma_{i}) \otimes (\mathbbm{1} +\sigma_j ), \, \nonumber\\
 &P_2 := \frac{1}{4} (\mathbbm{1} - \sigma_i) \otimes (\mathbbm{1} - \sigma_j), \, \nonumber\\
&  P_3 := \frac{1}{4} (\mathbbm{1} + \sigma_i) \otimes (\mathbbm{1} - \sigma_j), \, \nonumber\\
& P_4 :=\frac{1}{4} (\mathbbm{1} - \sigma_i) \otimes (\mathbbm{1} + \sigma_j),
\end{align}
are the elements of the projective measurement. The observable can also be decomposed in terms of the Bell basis:
\begin{align}\label{eq: statistics2}
 \sigma_i \otimes \sigma_j  &=   \tilde{P}_1   +    \tilde{P}_2  -     \tilde{P}_3  -  \tilde{P}_4  \nonumber\\
 &=   (\tilde{P}_1+\tilde{P}_2 ) -  (\tilde{P}_3 +\tilde{P}_4 )  ,
\end{align}
where 
\begin{align}
  \tilde{P}_1 & := \frac{1}{4}  U (\mathbbm{1}  \otimes \mathbbm{1} + \sigma_1 \otimes \sigma_1 - \sigma_2 \otimes \sigma_2+ \sigma_3 \otimes \sigma_3)  U^\dag ,  \nonumber\\
  \tilde{P}_2 & := \frac{1}{4}U (\mathbbm{1}  \otimes \mathbbm{1} + \sigma_1 \otimes \sigma_1 + \sigma_2 \otimes \sigma_2- \sigma_3 \otimes \sigma_3)  U^\dag , \nonumber\\
   \tilde{P}_3 & := \frac{1}{4} U(\mathbbm{1}  \otimes \mathbbm{1} - \sigma_1 \otimes \sigma_1 + \sigma_2 \otimes \sigma_2+ \sigma_3 \otimes \sigma_3)  U^\dag , \nonumber\\
  \tilde{P}_4 & :=\frac{1}{4}U (\mathbbm{1}  \otimes \mathbbm{1} - \sigma_1 \otimes \sigma_1 - \sigma_2 \otimes \sigma_2- \sigma_3 \otimes \sigma_3)  U^\dag ,
\end{align}
are elements of the Bell measurement with $U$ being any unitary satisfying $ U\left( \sigma_1 \otimes \sigma_1\right) U^\dag =  \sigma_i \otimes \sigma_j.$

One can show
\begin{align}\label{eq:P+}
  P_1   +   P_2   =   \tilde{P}_1+\tilde{P}_2=: P_+
\end{align}
and 
\begin{align}\label{eq:P-}
  P_3   +   P_4   =   \tilde{P}_3+\tilde{P}_4=: P_-.
\end{align}
We shall define the PDM in terms of the corresponding coarse-grained measurement
 $$ \Bigl\{P_+:=   \frac{  \mathbbm{1} \otimes \mathbbm{1} + \sigma_i \otimes \sigma_j  }{2}  ,  P_- :=    \frac{  \mathbbm{1} \otimes \mathbbm{1} - \sigma_i \otimes \sigma_j  }{2} \Bigl  \}.$$
 One possible way to implement the coarse-grained measurements is provided in the Supplementary Information.

\noindent {\bf {\em Note added.}}  The above quantum causal inference protocol has, after the preparation of this manuscript, been implemented experimentally in an NMR platform~\cite{liu2024experimental, liu2024quantumcausal}.

\noindent {\bf {\em Code Availability.}}  Codes are available upon request to the authors. 

\noindent {\bf {\em Data Availability.}}  No data was generated in this research apart from that presented in the paper.

\noindent {\bf {\em Acknowledgements.}} 
We thank Dong Yang, Daniel Ebler, Qian Chen, Caslav Brukner, Giulio Chiribella and James Fullwood for discussions. XL and OD acknowledge support from the NSFC (Grants No. 12050410246, No. 1200509, No. 12050410245) and City University of Hong Kong (Project No. 9610623). Part of this work was carried out when XL was visiting City University of Hong Kong. XL also acknowledge support from the National Research Foundation, Prime Minister’s Office, Singapore under its Campus for Research Excellence and Technological Enterprise (CREATE) programme. YQ is supported by the NRF, Singapore and A*STAR. This publication was made possible in part through the support of the ID 61466 grant from the John Templeton Foundation, as part of the The Quantum Information Structure of Spacetime (QISS) Project (qiss.fr). The opinions expressed in this publication are those of the authors and do not necessarily reflect the views of the John Templeton Foundation. This research is also funded in part by the Gordon and Betty Moore Foundation through Grant GBMF10604 to VV. 

\noindent {\bf {\em Author Contributions.}}
X.L. conceived the idea and derived the main results. All authors contributed to discussions throughout and the development of the results. X.L. and O.D. wrote the manuscript with inputs from all authors.  O.D. supervised the project.

\noindent {\bf {\em  Competing Interests.}} The authors declare no competing interests.

  \bibliography{ref}
 
  \newpage 

\onecolumngrid

\cleardoublepage
\setcounter{page}{1}
\setcounter{footnote}{0}
\thispagestyle{empty}

\begin{center}
	\textbf{\large Supplementary Information for Quantum Causal Inference with Extremely Light Touch}\\
	 \vspace{2ex}
	\text{Xiangjing Liu$^{1,2,3,4,5}$, Yixian Qiu$^{5}$, Oscar Dahlsten$^{3,6,7}$ and Vlatko Vedral$^{8}$ }\\
		 \vspace{1ex}
\textit{ $^1$ CNRS@CREATE, 1 Create Way, 08-01 Create Tower, Singapore 138602, Singapore}\\
\textit{ $^2$ MajuLab, CNRS-UCA-SU-NUS-NTU International Joint Research Unit, Singapore}\\
\textit{ $^3$ Department of Physics, City University of Hong Kong, 83 Tat Chee Avenue, Kowloon, Hong Kong}\\
\textit{ $^4$ Department of Physics, Southern University of Science and Technology, Shenzhen 518055, China}\\
\textit{$^5$ Centre for Quantum Technologies, National University of Singapore, Singapore 117543, Singapore}\\
\textit{$^6$ Shenzhen Institute for Quantum Science and Engineering, Southern University of Science and Technology, Shenzhen 518055, China}\\
\textit{$^7$ Institute of Nanoscience and Applications, Southern University of Science and Technology, Shenzhen 518055, China}\\
\textit{$^8$ Clarendon Laboratory, University of Oxford, Parks Road, Oxford OX1 3PU, United Kingdom}

\end{center}

\section{A method for implementing the coarse-grained measurement}
We now describe a method for implementing the coarse grained measurements $P_+$  and $P_-$ associated with the closed form expression of the PDM. As described in the Methods, for a given time point $$ \Bigl\{P_+:=   \frac{  \mathbbm{1} \otimes \mathbbm{1} + \sigma_i \otimes \sigma_j  }{2}  ,  P_- :=    \frac{  \mathbbm{1} \otimes \mathbbm{1} - \sigma_i \otimes \sigma_j  }{2} \Bigl  \}.$$

\begin{figure*}[h]
  \centering
  \includegraphics[scale=0.35]{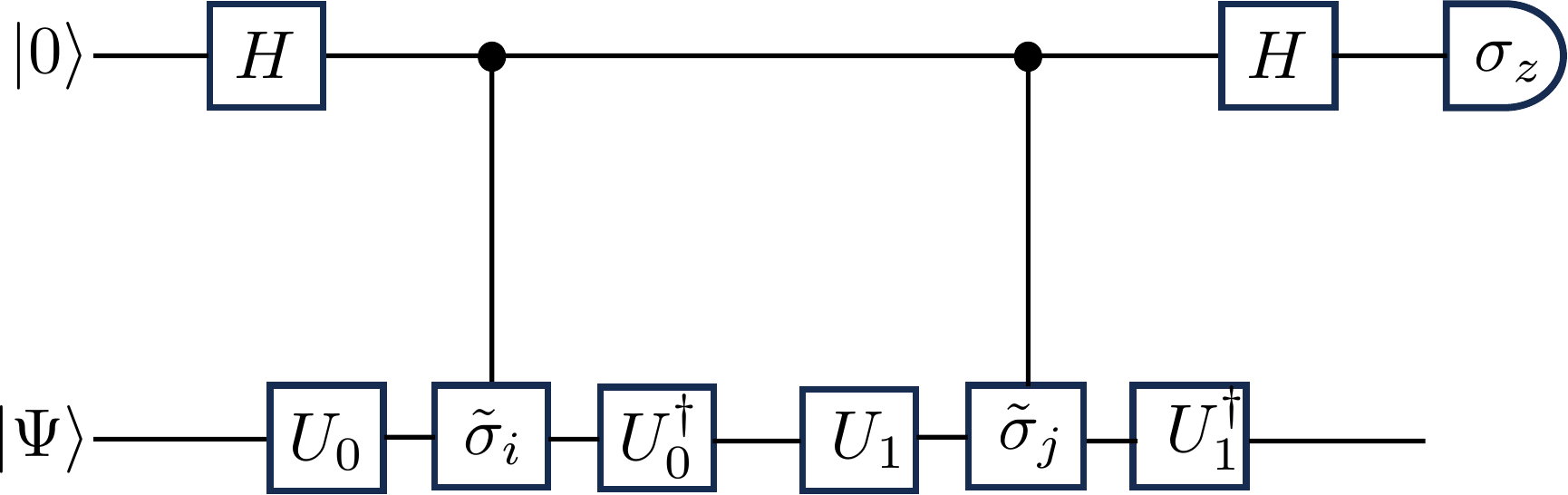}
  \caption{A quantum circuit for measuring time correlation function $ \langle \tilde{\sigma}_j(t_1)  \tilde{\sigma}_i (t_0) \rangle $.  $\sigma_3=\sigma_z$ and $\tilde{\sigma}_i$ is the $i$-th Pauli group element on $n$ qubits.  }  
  \label{fig:SC}
\end{figure*}

 One possible way of obtaining the statistics by coarse-grained measurements is via a quantum scattering circuit~\cite{souza2011scattering}. As depicted in FIG.~\ref{fig:SC}, the scattering circuit  takes the 
input state as $ \rho = \rho_{\text{ancilla}} \otimes \rho_s= \ket{0}\bra{0} \otimes \ket{\psi} \bra{\psi} $. One can calculate that the expectation value $ \langle \sigma_3 \rangle$ on the ancilla  equals to the two-point correlation function of the system:
\begin{align}\label{eq: twotimecorr}
 \langle \sigma_3 \rangle =  \langle \tilde{\sigma}_j (t_1) , \tilde{ \sigma}_i(t_0)  \rangle,
\end{align}
 where   $ \tilde{\sigma}_i (t_0):= U_0 \tilde{ \sigma}_i U_0^\dag := e^{-iht_0} \tilde{ \sigma}_i  e^{iht_0} $ and  $ \tilde{ \sigma}_j (t_1):= U_1 \tilde{\sigma}_j U_1^\dag := e^{-iht_1} \tilde{\sigma}_j  e^{iht_1} $.  
   Eq.~\eqref{eq: twotimecorr} can be rewritten as
 \begin{align} 
   \langle \tilde{\sigma}_j (t_1) , \tilde{ \sigma}_i(t_0)  \rangle =&   \langle  U_1 \ (P^j_+-P^j_-)  \ U_1^\dag  U_0 \ (P^i_+-P^i_-)  \ U_0^\dag  \rangle \nonumber\\
 =& \Bigl\{ \langle  U_1  P^j_+   \ U_1^\dag  U_0 P^i_+   \ U_0^\dag  \rangle +  \langle  U_1  P^j_-  \ U_1^\dag  U_0 P^i_-   \ U_0^\dag  \rangle \Bigl\}  \nonumber\\
&-  \Bigl\{ \langle  U_1  P^j_+   \ U_1^\dag  U_0 P^i_-   \ U_0^\dag  \rangle +  \langle  U_1  P^j_+ \ U_1^\dag  U_0 P^i_-   \ U_0^\dag  \rangle \Bigl\} .
\end{align}
 Thus, Eq.~\eqref{eq: twotimecorr} gives the same value as the expectation value obtained by coarse-grained measurements $P^i_+$  and $P^i_-$ taking place initially and finally.

\section{Derivation of the closed-form of 2-time $n$-qubit PDM}
 Having the quantum state $\rho_{1}$ at time $t_1$,  the measurement $\{ P^{\alpha}_+, P^{\alpha}_-\}$ at each time $t_\alpha, \alpha=1,2,$ and the quantum channel $\mathcal{M}_{2|1}$ with its CJ matrix $M_{12}$, we are in a good position to obtain the 2-time $n$-qubit PDM. Recall that the general form of an $n$-qubit PDM across two times $ [t_1, t_2] $ is given by 
\begin{align}
R_{12}=\frac{1}{2^{2n}} \sum^{4^n-1}_{i_1=0} \sum^{4^n-1}_{i_2=0}  \langle  \tilde{\sigma}_{i_{1}} ,  \tilde{\sigma}_{i_{2}} \rangle \,  \tilde{\sigma}_{i_1} \otimes  \tilde{\sigma}_{i_2} .
\end{align}
Therefore, in order to construct the PDM, it is necessary to calculate the expectation values of the product of the result of the measurements, i.e., $ \langle  \tilde{\sigma}_{i_{1}} ,  \tilde{\sigma}_{i_2} \rangle$. According to the definition of the PDM, the initial state collapse to the eigenstates of the $n$-qubit Pauli matrix $\tilde{\sigma}_{i_{1}}$ after the measurement at time $t_1$. Then the post-measurement states go through the channel followed by a measurement at time $t_2$. Denoting $\{ P^{\alpha}_{+}=  (  \mathbbm{1} + \tilde{ \sigma}_{i_\alpha } )/2, P^{\alpha}_{-}= (  \mathbbm{1} - \tilde{ \sigma}_{i_\alpha} )/2   \}$ as the coarse-grained measurement scheme at time $t_\alpha, \alpha=1,2$,  the product of expectation values can be expressed by
\begin{align}\label{eq: expectations}
& \langle  \tilde{\sigma}_{i_{1}} ,  \tilde{\sigma}_{i_2} \rangle  = \Tr[M_{12}  (P^1_+ \rho_1 P^1_+ \otimes  \tilde{\sigma}_{i_2})  ] -\Tr[M_{12}  (P^1_- \rho_1 P^1_- \otimes  \tilde{\sigma}_{i_2})  ], 
\end{align}
The equality, $\mathcal{M}_{2|1} (\rho_1) = \Tr_1  [M_{12} \rho_1 \otimes \id_2] $, which holds for our choice of convention for the CJ matrix, will be frequently used in the calculations.
It is straightforward to calculate that 
 \begin{align}\label{eq: poststates}
  P^1_+ \rho_1 P^1_+ - P^1_- \rho_1 P^1_-
   =   \frac{  \mathbbm{1} + \tilde{ \sigma}_{i_1}  }{2}  \rho_1 \frac{  \mathbbm{1} + \tilde{ \sigma}_{i_1}  }{2}   -  \frac{  \mathbbm{1} -  \tilde{ \sigma}_{i_1}  }{2}   \rho_1 \frac{  \mathbbm{1} - \tilde{ \sigma}_{i_1}  }{2} 
 = \frac{1}{2} ( \rho_1  \tilde{ \sigma}_{i_1}  +\tilde{ \sigma}_{i_1} \rho_1  ).
 \end{align}
 Substituting Eq~\eqref{eq: poststates} into Eq~\eqref{eq: expectations} leads to
\begin{align} \label{eq: ExpValue}
  \langle  \tilde{\sigma}_{i_{1}} ,  \tilde{\sigma}_{i_2} \rangle  
= &\Tr [ \frac{1}{2} M_{12} (\rho_1  \tilde{ \sigma}_{i_1}  +\tilde{ \sigma}_{i_1} \rho_1 ) \otimes  \tilde{\sigma}_{i_2} ]\nonumber\\
=&\Tr [ \frac{1}{2} (M_{12} \, \rho_1 \otimes \mathbbm{1}_2  +  \rho_1 \otimes \mathbbm{1}_2 \, M_{12} )  \tilde{\sigma}_{i_{1}} \otimes  \tilde{\sigma}_{i_2} ] \nonumber\\
 \equiv & \Tr [ R_{12} \tilde{\sigma}_{i_{1}}  \otimes \tilde{\sigma}_{i_2} ].
\end{align} 
where the cyclic property of the trace is used in the second equality. Finally, the two-qubit PDM is expressed by 
\begin{align}\label{eq: two-qPDM}
R_{12} = \frac{1}{2}( M_{12} \, \rho +\rho \, M_{12}  ) .
\end{align}
where $\rho := \rho_1 \otimes \mathbbm{1}_2 $.

In the following, we verify that the PDM given by Eq.~\eqref{eq: two-qPDM} returns to the normal density matrix when there is one time point left.
\begin{itemize}

\item[1)] Tracing out the final time $t_2$, the PDM returns to the initial state,
 \begin{align}
\Tr_2 R_{12} = \frac{1}{2} \Tr_2  ( M_{12} \, \rho +\rho \, M_{12}  ) = \rho_{1},
\end{align} where $\Tr_2 M_{12}= \mathbbm{1} _1$ is used in the second equality,

\item[2)] Tracing out the initial time $t_1$, the PDM returns to the output state,
\begin{align} \label{eq: finalstate}
 \Tr_1 R_{12} = \frac{1}{2} \Tr_1  ( M_{12} \, \rho +\rho \, M_{12}  ) = \Tr_1  M_{12} \, \rho_{1} \otimes \mathbbm{1}_2 = \mathcal{M}_{2|1}(\rho_{1}) ,
 \end{align}
 where the cyclic property of the trace is used in the second equality.

\end{itemize}

\section{Derivation of the closed-form of $m$-time $n$-qubit PDM}

 In this section, we extend the $n$-qubit PDM formalism to across $m$ times. To do that we first consider an $n$-qubit state $\rho$ at time $t_1$ that undergoes the channel $\mathcal{M}_{2|1}$,  arrives at time $ t_2$, then experiences channel $\mathcal{M}_{3|2}$ and arrives at time $t_3$. Denote the CJ matrices for channels $\mathcal{M}_{2|1}, \mathcal{M}_{3|2}$ by $M_{12}, M_{23}$ respectively. The tensor product `$ \otimes$' is sometimes omitted when there is no confusion. 

We first recall that the two-time $n$-qubit PDM takes the form
\begin{align}
R_{12}= \frac{1}{2} (M_{12} \rho + \rho M_{12} ).
\end{align}

We now proceed to derive a closed form of the 3-time $n$-qubit PDM. To do that,  it is sufficient to calculate the expectation values of measurement events. Consider measurement events at times $t_1,t_2, t_3$ being $ \{ \tilde{\sigma}_{i_1}\}, \{ \tilde{\sigma}_{i_2} \}, \{ \tilde{\sigma}_{i_3} \}$, respectively. The initial state $\rho_1$ first collapses to the eigenstates of operator $ \tilde{\sigma}_{i_1}$ then goes into the channel $\mathcal{M}_{2|1}$. The central quantity we wish to evaluate is $\langle  \tilde{\sigma}_{i_1} , \tilde{ \sigma}_{i_2} ,\tilde{ \sigma}_{i_3} \rangle$.

Our expression for $\langle  \tilde{\sigma}_{i_1} , \tilde{ \sigma}_{i_2} ,\tilde{ \sigma}_{i_3} \rangle$ will turn out to involve a matrix $\rho_2^{( \tilde{\sigma}_{i_1}) }$. We now derive a simplified expression for $\rho_2^{( \tilde{\sigma}_{i_1}) }$, which we shall use later.
\begin{align}\label{eq: rho2}
\rho_2^{( \tilde{\sigma}_{i_1}) } :=& \Tr_1 [ \frac{1}{2} (M_{12} P^{1}_{ +} \rho_1 P^{1}_{ +}  +   P^{1}_{ +}  \rho_1   P^{1}_{ +}  M_{12} ) ]- \Tr_1 [ \frac{1}{2} (M_{12}  P^{1}_-  \rho_1   P^{1}_{ -}  +   P^{1}_{ -}  \rho_1   P^{1}_-  M_{12} )  ]\nonumber\\
=&  \Tr_1 [ \frac{1}{2} M_{12} ( P^{1}_{ +} \rho_1 P^{1}_{ +}  -  P^{1}_-  \rho_1   P^{1}_{ -}   ) ]+ \Tr_1 [ \frac{1}{2} ( P^{1}_+  \rho_1   P^{1}_{ +}  -   P^{1}_{ -}  \rho_1   P^{1}_- ) M_{12} ]  \nonumber\\
=&  \Tr_1 [ \frac{1}{4} M_{12}  ( \rho_1  \tilde{ \sigma}_{i_1}  +\tilde{ \sigma}_{i_1} \rho_1  )]
 + \Tr_1 [ \frac{1}{4}  ( \rho_1  \tilde{ \sigma}_{i_1}  +\tilde{ \sigma}_{i_1} \rho_1  ) M_{12}  ] \nonumber\\
 =&  \Tr_1 [ \frac{1}{2}   ( M_{12} \rho_1    + \rho_1 M_{12} )\tilde{ \sigma}_{i_1} ]  \nonumber\\
 =&\Tr_1 [ R_{12} \tilde{\sigma}_{i_1} ],
\end{align}
where the identity operators $\mathbbm{1}_2$ at time $t_2$ are ignored throughout the calculation, Eq.~\eqref{eq: poststates} is used in the third equality and the cyclic property of the trace is used in the fourth equality. Then, the expectation value of the product of measurement outcomes at three time points is given by
\begin{align}
 & \langle  \tilde{\sigma}_{i_1} , \tilde{ \sigma}_{i_2} ,\tilde{ \sigma}_{i_3} \rangle \nonumber\\
 =&\Tr_{23}  [ M_{23} \Bigl( P^{2}_{ +} \rho^{ (\tilde{\sigma}_{i_1}) }_2 P^{2}_{+} - P^{2}_{-}\rho^{ (\tilde{\sigma}_{i_1}) }_2 P^{2}_{ -}  \Bigl)  \tilde{ \sigma}_{i_3} ] \nonumber\\
 =& \frac{1}{2} \Tr_{23}  [ M_{23} \Bigl(  ( \Tr_{1}  R_{12} \tilde{\sigma}_{i_1}  ) \sigma_{i_2}+ \tilde{\sigma}_{i_2} ( \Tr_{1}  R_{12} \tilde{ \sigma}_{i_1} )   \Bigl)  \tilde{\sigma}_{i_3} ] \nonumber\\
 =& \frac{1}{2} \Tr_{123}  [( M_{23}   R_{12} \tilde{ \sigma}_{i_1}  \tilde{\sigma}_{i_2} +    R_{12} \tilde{\sigma}_{i_1} M_{23}   \tilde{ \sigma}_{i_2} )    \tilde{\sigma}_{i_3} ] \nonumber\\
 =& \frac{1}{2} \Tr   [ ( M_{23}  R_{12}   +    R_{12}  M_{23}  ) \tilde{ \sigma}_{i_1}  \tilde{ \sigma}_{i_2}  \tilde{\sigma}_{i_3} ] ,
\end{align} 
where Eq.~\eqref{eq: poststates} and Eq.~\eqref{eq: rho2} are used in the second equality and the cyclic property of the trace is used in the third equality.
Thus, we have the $n$-qubit PDM across 3 times as 
\begin{align}
R_{123}=\frac{1}{2}  ( R_{12}   M_{23} +   M_{23} R_{12} ).
\end{align}

Let us assume that we have the  $(m-1)$-time $n$-qubit PDM $R_{12,...,m-1}$. We then consider $m$ events $\{ \tilde{\sigma}_{i_1} , \tilde{\sigma}_{i_2}, ..., \tilde{\sigma}_{i_{m}}    \}$ and the channel $\mathcal{M}_{i+1|i}$ between time interval $[t_{i}, t_{i+1}], i\in \{1,2,3 ,...,m-1\}$ with the corresponding CJ matrix $M_{i,i+1}$. Similarly, the matrix $\rho^{(\vec{ \tilde{\sigma} }_{} )}_{m-1}$  just before measurement made at time $t_{m-1}$ can be written as 
\begin{align}
\rho^{(\vec{ \tilde{\sigma} }_{} )}_{m-1}= \Tr_{12...m-2} [ R_{12...m-1} \tilde{\sigma}_{i_1} \tilde{\sigma}_{i_2}...\tilde{\sigma}_{i_{m-2}} ].
\end{align}
The expectation value of the product of measurement outcomes at those time points is given by
\begin{align}
 & \langle  \tilde{\sigma}_{i_1} , \tilde{ \sigma}_{i_2} ,...,\tilde{ \sigma}_{i_{m}} \rangle \nonumber\\
 =&\Tr_{m-1,m}  \Bigl( M_{m-1,m} \Bigl( P^{m-1}_{ +} \rho^{(\vec{ \tilde{\sigma} }_{} )}_{m-1} P^{m-1}_{+} - P^{m-1}_{-}\rho^{(\vec{ \tilde{\sigma} }_{} )}_{m-1} P^{m-1}_{ -}  \Bigl)  \tilde{ \sigma}_{i_m} \Bigl) \nonumber\\
 =& \frac{1}{2} \Tr_{m-1,m}  \Bigl( M_{m-1,m} \Bigl(  ( \Tr_{12...m-2}  R_{12...m-1}  \tilde{\sigma}_{i_1} \tilde{\sigma}_{i_2} ...\tilde{\sigma}_{i_{m-2}}  ) \tilde{\sigma}_{i_{m-1}}  \nonumber\\
 &+ \tilde{\sigma}_{i_{m-1}}( \Tr_{12...m-2}  R_{12...m-1}  \tilde{\sigma}_{i_1} \tilde{\sigma}_{i_2} ...\tilde{\sigma}_{i_{m-2}}  )   \Bigl)  \tilde{\sigma}_{i_m} \Bigl) \nonumber\\
 =& \frac{1}{2} \Tr_{12...m}  \Bigl( \Bigl( M_{m-1,m} ( R_{12...m-1}  \tilde{\sigma}_{i_1} \tilde{\sigma}_{i_2} ...\tilde{\sigma}_{i_{m-2}}   ) \tilde{\sigma}_{i_{m-1}}  \nonumber\\
 & +  ( R_{12,...,m-1}  \tilde{\sigma}_{i_1} \tilde{\sigma}_{i_2} ...\tilde{\sigma}_{i_{m-2}}  ) M_{m-1,m}   \tilde{ \sigma}_{i_{m-1}}    \Bigl) \tilde{\sigma}_{i_m} \Bigl) \nonumber\\
 =& \frac{1}{2} \Tr_{12...m}   \Bigl( ( M_{m-1,m}  R_{12...m-1}   +    R_{12...m-1}  M_{m-1,m}  ) \tilde{ \sigma}_{i_1} \tilde{ \sigma}_{i_2} ... \tilde{\sigma}_{i_m} \Bigl) ,
\end{align} 
Thus, we have $R_{12...m}= \frac{1}{2} (R_{12...m-1} M_{m-1,m}+  M_{m-1,m} R_{12...m-1}  ) $.

In summary,
the $n$-qubit PDM across $m$ times is given by the following iterative expression
\begin{align}
R_{12...m}= \frac{1}{2} (R_{12...m-1} M_{m-1,m}+  M_{m-1,m} R_{12...m-1}  )
\end{align}
with the initial condition $R_{12}= \frac{1}{2}( \rho  M_{12} +M_{12} \rho  ) $ and with $M_{m-1,m}$ denoting the CJ matrix of the $(m-1)$th channel.

 \section{ Partial trace property of PDM} 
 Given a pseudo-density matrix $R_{A_1B_1A_2B_2}$ defined over two sets of events $A_1B_1$ and $A_2B_2$ and a pseudo-density matrix $R_{A_1B_2}$ constructed from the sets of events $A_1$ and $B_2$,  one can show that the latter can be obtained from $R_{A_1B_1A_2B_2}$ by tracing over the subsystem corresponding to $B_1A_2$, i.e., $ R_{A_1B_2} = \Tr_{A_2B_1}R_{A_1B_1A_2B_2} $. 
  \begin{proof}
   For the simplicity of the notations, we give a proof for the two-time two-qubit PDM case, which can be directly generalized to the $m$-time $n$-qubit PDM case.  In the following calculation, the tensor product `$ \otimes$' is omitted when there is no risk of confusion.
  
   Based on the definition of pseudo-density matrix $R_{A_1A_2B_1B_2}$, we have
      \begin{align}
  R_{A_1B_1A_2B_2}= \frac{1}{2^4} \sum^3_{i,j,k,l=0} \langle \sigma^{A_1}_i \sigma^{B_1}_j, \sigma^{A_2}_k \sigma^{B_2}_l  \rangle \sigma^{A_1}_i \sigma^{B_1}_j \sigma^{A_2}_k \sigma^{B_2}_l.
    \end{align}
    We define
  \begin{align}
  R_{A_1B_2}= \frac{1}{4 } \sum^3_{i=0} \sum^3_{j=0} \langle \sigma^{A_1}_i \sigma_0^{B_1} , \sigma^{A_1}_0 \sigma^{B_2}_j  \rangle \sigma^{A_1}_i  \sigma^{B_2}_j , \nonumber
    \end{align}
  which we express using shorthand as
  \begin{align}
  R_{A_1B_2}= \frac{1}{4 } \sum^3_{i=0} \sum^3_{j=0} \langle \sigma^{A_1}_i ,  \sigma^{B_2}_j  \rangle \sigma^{A_1}_i  \sigma^{B_2}_j .
    \end{align}

It is then straightforward to calculate that
\begin{align}
& \Tr_{B_1A_2}   R_{A_1B_1A_2B_2} \nonumber\\
=&  \Tr_{B_1A_2} [ \frac{1}{2^4} \sum^3_{i,j,k,l=0} \langle \sigma^{A_1}_i \sigma^{B_1}_j, \sigma^{A_2}_k \sigma^{B_2}_l  \rangle \sigma^{A_1}_i \sigma^{B_1}_j \sigma^{A_2}_k \sigma^{B_2}_l ] \nonumber\\
=&  \frac{1}{2^4} \sum^3_{i,j,k,l=0}  \Tr_{B_1A_2}(\sigma^{B_1}_j   \sigma^{A_2}_k)   \langle \sigma^{A_1}_i \sigma^{B_1}_j, \sigma^{A_2}_k \sigma^{B_2}_l  \rangle \sigma^{A_1}_i \sigma^{B_2}_l  \nonumber\\
=&  \frac{1}{2^2} \sum^3_{i,j=0}    \langle \sigma^{A_1}_i \sigma^{B_1}_0, \sigma^{A_2}_0 \sigma^{B_2}_j  \rangle \sigma^{A_1}_i \sigma^{B_2}_j  \nonumber\\
=&R_{A_1B_2}.
\end{align}
This completes the proof.
\end{proof}

\section{Proof of Theorem~\ref{thm:nullneg} }

Theorem~\ref{thm:nullneg}.\textit{
If a quantum channel $\mathcal{P}$ does not allow signaling from $B$ to $A$, then, for any state $\rho_{A_1B_1}$ at time $t_1$,  the PDM $R_{B_1A_2} $ is positive semidefinite and the PDM negativity $f(R_{B_1A_2} )=0$. 
}

\begin{proof} 

A channel $\mathcal{P}: \mathcal{B}(\mathcal{H}_{A_1B_1}) \rightarrow   \mathcal{B}(\mathcal{H}_{A_2B_2}) $ is called semicausal~\cite{eggeling2002semicausal}, if, 
\begin{align} \label{eq: semicausal}
\Tr_{B_2} \mathcal{P}(\rho_{A_1B_1}) = \mathcal{T}  (\rho_{A_1}),
\end{align}
for all input states $\rho_{A_1B_1}$ and for some completely positive maps $ \mathcal{T}: \mathcal{B}(\mathcal{H}_{A_1}) \rightarrow   \mathcal{B}(\mathcal{H}_{A_2}) $. (Ref.~\cite{eggeling2002semicausal} moreover shows that semicausal is equivalent to semilocalizable )

   For the simplicity of the notations, we give a proof to the two-qubit PDM case, which can be directly generalized to the $n$-qubit PDM case. In the Pauli basis, the CJ matrix of channel $\mathcal{P}$ is 
   \begin{align} \label{eq: pChoi}
 M= &\sum^{2^2-1}_{i ,j=0}\ket{i} \bra{j}^T \otimes  \mathcal{P} \left( \ket{i}\bra{j}  \right) = \frac{1}{2^2} \sum^3_{i,j=0} \sigma^{A_1}_i \sigma^{B_1}_j \mathcal{P} ( \sigma^{A_1}_i \sigma^{B_1}_j     ) .
 \end{align}
 where the tensor products `$\otimes$' are omitted.
   The PDM can be calculated as 
\begin{align}
  R_{B_1A_2}  =& \Tr_{A_1B_2} R_{A_1B_1A_2B_2}  \nonumber\\
  =& \frac{1}{2}  \Tr_{A_1B_2} [ M  \rho_{A_1B_1} +\rho_{A_1B_1} M  ] \nonumber\\
  =&  \frac{1}{8}  \Tr_{A_1B_2} \sum^3_{i,j=0} \{ \rho_{A_1B_1}, \sigma^{A_1}_i \sigma^{B_1}_j \} \mathcal{P} ( \sigma^{A_1}_i \sigma^{B_1}_j     )  \nonumber\\
  =&   \frac{1}{8} \Tr_{A_1}\sum^3_{i=0} \{ \rho_{A_1B_1}, \sigma^{A_1}_i  \sigma^{B_1}_0 \}   \mathcal{T}( \sigma^{A_1}_{i} )  \nonumber\\
  =&  \frac{1}{4}  \sum^3_{i,l=0} \langle \sigma^{A_1}_i  \sigma^{B_1}_l \rangle \sigma^{B_1}_l   \mathcal{T}( \sigma^{A_1}_{i} )  \nonumber\\
  =&   \id_{ B} \otimes \mathcal{T} ( S \rho_{A_1B_1} S^\dag)   , 
\end{align}
where $S$ is the swap operator, $\{ \,,\}$ denotes the anticommutator,  Eq.~\eqref{eq: pChoi} is used in the third equality and Eq.~\eqref{eq: semicausal} is used in the fourth equality.
Since $\mathcal{T}$ is completely positive, it is straightforward to see that $ R_{B_1A_2}$ is positive semidefinite. Finally, $f(R_{B_1A_2})=0$ since the causal monotone $f$ measures the negativity of PDMs. This completes the proof.
\end{proof}

The following Corollary follows immediately from Theorem~\ref{thm:nullneg}.
\begin{cor}
If a quantum channel $\mathcal{P}$ does not allow signaling in both directions, then, for any state $\rho_{A_1B_1}$ at time $t_1$, the two PDMs $R_{B_1A_2}$, $R_{A_1B_2} $ are positive semidefinite and the PDM negativities $f(R_{B_1A_2} )=f(R_{A_1B_2} )=0$.
\end{cor}

\section{Determining causal influence of channel by optimizing the input state} 
We conjecture that, for the case where the input state is free to choose because one is solely interested in whether the channel is signaling, there is always a family of states such that $f >0$ provides a necessary and sufficient condition for the channel to allow causal influence in a given direction. We here show the conjecture to hold for a quite general class of two-qubit unitary channels. 

We slightly restrict the general two-qubit unitary $U_{AB}$ decomposition from Ref.~\cite{kraus2001optimal}, and demand  
\begin{align}
U_{AB}=U_A \otimes U_B \, e^{i\theta S_{AB}} \,V_A \otimes V_B,
\end{align}
where $S$ denotes the swap operator, and  the operator $e^{i\theta S_{AB}}$ is the reason for causal influence from both sides.  By property (II), the PDM negativity $f(R)$ is invariant under local change of basis. Thus one can ignore the two factors $U_A \otimes U_B$ and $V_A \otimes V_B$ in $U_{AB}$ when analysing causal influence via $f$. 

We now provide an example to illustrate that an input state can be found to witness causal influence of the unitary operator $e^{i\theta S_{AB}}$. Consider the state of the compound system $\rho_{A_1B_1}= \ket{0}_{A_1}\bra{0} \otimes \ket{0}_{B_1}\bra{0}$ undergoing a unitary evolution:
$$
U_{AB}(\theta)=e^{i \theta S}= \cos(\theta) \,  \id + i  \sin(\theta) \, S := c \id + i s S.
$$
Under the evolution $U_{AB}(\theta)$, the initial state $\rho_{A_1}$ of the system cannot influence its final state $\rho_{A_2}$ when $c=0$, i.e., there is no causal influence from $A_1$ to $A_2$. Moreover, there is causal influence from $A_1$ to  $A_2$ when $c \neq 0$.  Similar analysis is also applicable to examining causal influences from   $A_1$ to $B_2$, $B_1$ to $B_2$, and $B_1$ to $A_2$.

The effective dynamics of the system $A$ can be modeled as a quantum channel with the set of Kraus operators given by 
$ \{ c \id + i  s \ket{0} \bra{0},  is \ket{0}\bra{1}  \} $ ~\cite{liu2022thermodynamics}.
The corresponding 2-time PDM of the system $A$ is given by 
\begin{align}
R_{A_1A_2} = \ket{00}\bra{00}+  \frac{c(c + i s)}{2} \ket{01}\bra{10} +  \frac{c(c - i s)}{2} \ket{10}\bra{01}.
\end{align}
The eigenvalues of $R_{A_1A_2}$ are $\{ 1, - \frac{c}{2},\frac{c}{2} , 0\}$.  Therefore, $f(R_{A_1A_2}  ) =|c| $. Therefore, under the choice of the initial state $\rho_{A_1B_1}= \ket{0}_{A_1}\bra{0} \otimes \ket{0}_{B_1}\bra{0}$,  $f(R_{A_1A_2}  ) >0$ provides a necessary and sufficient condition for the parameterized channel to allow causal influence from $A_1$ to $A_2$.

\section {Pure cause-effect mechanism is CP}
 Suppose that the pure cause-effect mechanism shown in case 1 of FIG.~\ref{fig:qsetup} is represented by no initial correlation and the semi-causal channel $\mathcal{P} = \mathcal{M}_{BC} \circ \mathcal{N}_{AC}$. 
$C$ is the intermediate system in the semi-localizable channel. We aim to show that the process $\map L _{A \rightarrow B}$ induced by $\mathcal{P}$ is a CP map.

First, suppose that the channel $\map P$ admits the unitary dilation $U$ such that
\begin{align}
\map P (\rho_{AB}) = \Tr_{\textcolor{red}{C}E}  \left( U (\rho_{AB} \otimes \rho_C \otimes \rho_E  )U^\dag  \right),
\end{align}
 where $E$ is used for the unitary dilation of $\mathcal{P}$.
Second, there is no initial correlation between $A$ and $B$, meaning $\rho_{AB}=\rho_A \otimes \rho_B$ where $\rho_A=\Tr_B \rho_{AB} $ and  $\rho_B=\Tr_A \rho_{AB}$.
Combining the two points, the process  $ \map L _{A \rightarrow B}$ is given by 
\begin{align}
\label{eq:forwardCP}
\map L _{A \rightarrow B} (\rho_{A}) = \Tr_{AC} \map P (\rho_{AB})  = \Tr_{ACE}  \left( U (\rho_A \otimes \rho_B \otimes \rho_C \otimes \rho_E  )U^\dag  \right).
\end{align}
It is not hard to see that the action of $\map L _{A \rightarrow B} $ can be expressed as the sum of Kraus operators acting on the state. Therefore, the process $\map L _{A \rightarrow B} $ is CP.

\section{Supporting details for quantum causal inference} 
Here we provide detailed reasoning and techniques for the proposed quantum causal inference protocol.

\begin{figure*}
  \centering
  \includegraphics[width=0.9\textwidth]{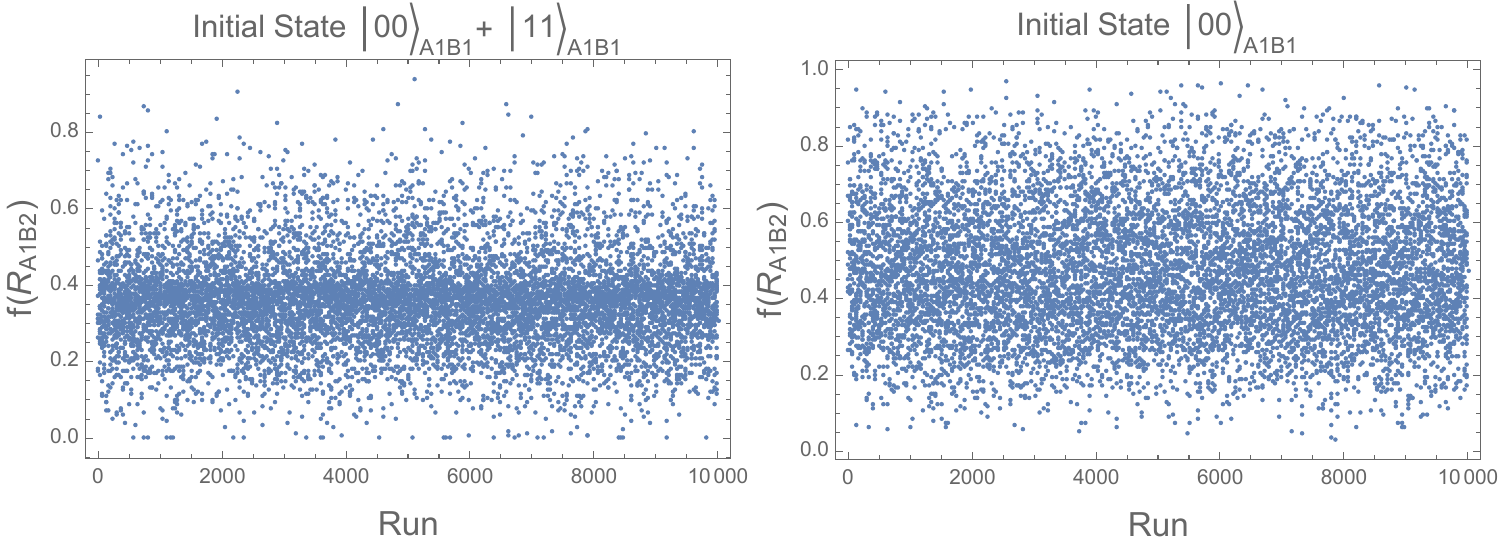}
  \caption{ {\bf Simulations of negativity of PDM with fixed input and random circuits.}   We set the channel $\mathcal{N}$ in FIG.~\ref{fig:semicausal} to be  $\mathcal{N}( \cdot )=S(\cdot) S^\dag $. $\mathcal{N}$ is here modelled as a unitary channel from $AC$ to $AC$ and $\mathcal{M}$ similarly from $BC$ to $BC$.  Then, we simulate the negativity $f(R_{A_1B_2})$ of the PDM $R_{A_1B_2}$ for two pure inputs, $ \psi_{A_1B_1}=\frac{1}{\sqrt{2}} (|00\rangle +|11\rangle )$ and $\psi_{A_1B_1}=|00\rangle$, with  $\mathcal{M}$ being  a Haar-random 2-qubit unitary channel in FIG.~\ref{fig:semicausal}. There is a cause-effect mechanism between $A$ and $B$  when $\mathcal{M}$ is not a local unitary channel.  We observe that almost all cases have negativity, therefore the negativity rules out the common cause mechanism with a near unit probability in the scenarios. }
  \label{fig:simulation1}
\end{figure*}

\begin{figure*}
  \centering
  \includegraphics[width=0.9\textwidth]{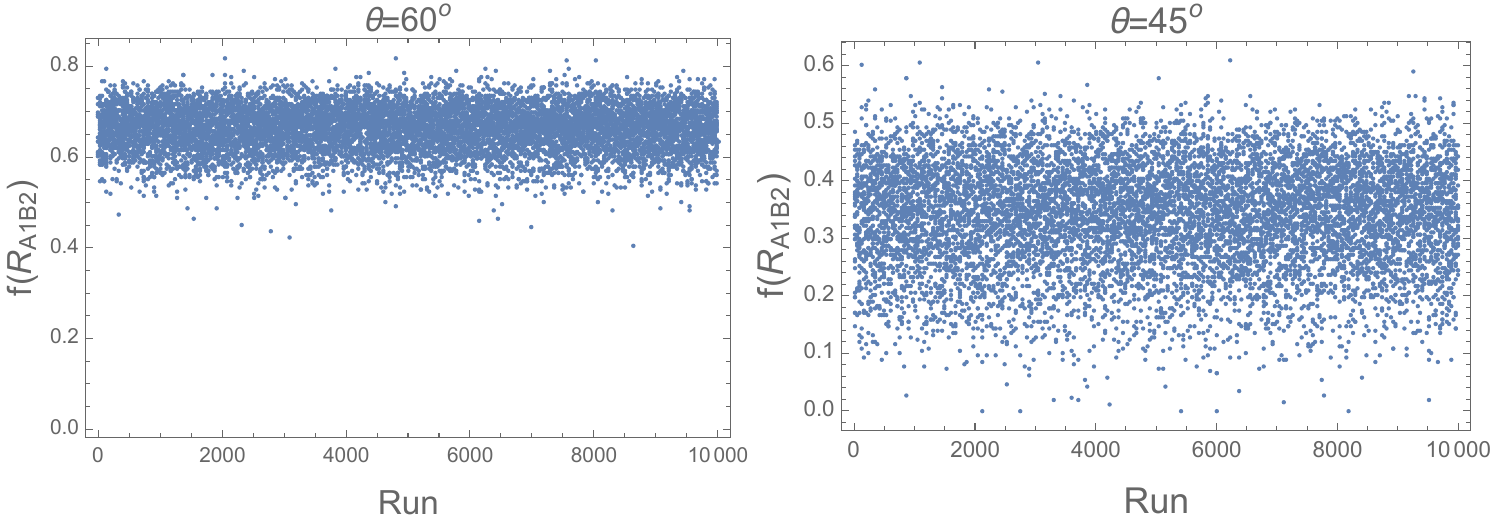}
  \caption{  {\bf Simulations of negativity of PDM with fixed circuit and random inputs.}  We set $\mathcal{N}( \cdot )=S(\cdot) S^\dag$ and $\mathcal{M}= U(\theta) ( \cdot ) U^\dag(\theta) $ with $U(\theta)=e^{-i \theta S} := \cos(\theta )\mathbb{1}-i \sin(\theta)S $. In this setting, there is a cause-effect mechanism between $A$ and $B$ when $\sin(\theta) \neq 0$. We simulate the negativity $f( R_{A_1B_2})$ of the PDM $R_{A_1B_2} $ for cases of $\theta=30^\circ $ and $\theta=60^\circ $ with the pure input state $|\psi\rangle_{A_1B_1}$ being Haar-random. We observe that almost all cases have negativity, therefore the negativity rules out the common cause mechanism with a near unit probability in the scenarios. }  
  \label{fig:simulation2}
\end{figure*}

Firstly, if $ f(R_{AB}) = 0$, then, according to Proposition~\ref{eq:RisaRho}, $R_{AB}$ can be treated as a density matrix, making it compatible with the common cause mechanism. However, this does not rule out the possibility of $R_{AB}$ representing other causal structures. Nevertheless, we numerically study the behaviour of the negativity of the PDM in FIGs.~\ref{fig:simulation1} and~\ref{fig:simulation2} and find that $f(R_{AB})>0$ with a near unit probability when there is a causal-effect mechanism between $A$ and $B$.
\begin{prop}
\label{eq:RisaRho}
Given that one only has access to the data that constructs a two-time PDM $R_{AB}$, if $f(R_{AB})=0$, then $R_{AB}$ is indistinguishable from the density matrix $\rho_{AB}$ that shares the same entries as $R_{AB}$.
\end{prop}
\begin{proof} The proposition is a result of Ref.~\cite{fitzsimons2015quantum}. According to the definition, any two-time PDM $R_{AB}$ is Hermitian and has a unit trace. These two properties, along with $f(R_{AB})= 0$, lead to the construction of a density matrix $\rho_{AB}$ with the same entries as $R_{AB}$. Given that we only have access to the data, there is no way to tell whether it is a density matrix or a two-time PDM.
\end{proof}


Secondly, if $f(R_{AB}) > 0$, there must exist temporal correlations, indicating the involvement of the cause-effect mechanism.  In this case, one needs the Choi matrix of the process and its time reversal to tell the cause and effect. To extract the CJ matrix $M$ from $R$, we employ the vectorization technique as described in Ref.~\cite{liu2023inferring}.
For any quantum operator 
$ A= \sum_{ij}  A_{ij} \ket{i}\bra{j} \in \mathcal{B}(H)  $, its vectorization is given by 
\begin{align}
 | A  \kk = \sum_{ij} A_{ij} \ket{i} \otimes \ket{j}  \in H \otimes H.
\end{align}
The vectorization of the PDM $R $ is then given by
\begin{align} 
| R  \kk =& \frac{1}{2} | \rho \, M+ M \, \rho  \kk   = \frac{1}{2} \left( \rho_1 \otimes \id_2 \otimes \id_1 \otimes \id_2 +  \id_1\otimes \id_2 \otimes \rho_1^T\otimes \id_2 \right)  | M \kk  =: B |M \kk,
\end{align}
where $ | EFG \kk=G \otimes E^T \, |F \kk$  is used in the second equality. When $\rho_1$ is full rank, then the complete information of $M$ is extracted via $|M \kk = B^{-1} |R \kk$. However, when $\rho_1$ is rank deficient, only the process information on the support of $B$ is obtained via $ |M \kk = B^{\ddag} |R \kk$ where $B^\ddag$ denotes the pseudo-inverse of $B$. 

Let us consider an example when $\rho_1$ is rank deficient. Suppose $\rho_1=\ket{0}\bra{0} $ undergoes an identity channel. The CJ matrix for the identity channel, $\rho(=\rho_1 \otimes \id_2) $ and the corresponding PDM are given by
\begin{align}
M =
    \begin{pmatrix}
        1 & 0  & 0  &  0  \\
        0 & 0  & 1  &  0  \\
        0 &  1 &0  &  0  \\
        0 &  0 &0  &  1  \\
    \end{pmatrix}, 
\, \,     \rho =
    \begin{pmatrix}
        1 & 0  & 0  &  0  \\
        0 & 1  & 0  &  0  \\
        0 &  0 &0  &  0  \\
        0 &  0 &0  &  0  \\
    \end{pmatrix}, 
\, \,    R= \frac{1}{2}(\rho M+M \rho) =
    \begin{pmatrix}
        1 & 0  & 0  &  0  \\
        0 & 0  & 1/2  &  0  \\
        0 &  1/2 &0  &  0  \\
        0 &  0 &0  &  0  \\
    \end{pmatrix}.
\end{align}
Working backwards to recover $M$ using $|N \kk = B^{\ddag} |R \kk$, where $N$ denotes the  recovered form for $M$, yields
\begin{align}
N =
    \begin{pmatrix}
        1 & 0  & 0  &  0  \\
        0 & 0  & 1  &  0  \\
        0 &  1 & a  &  b  \\
        0 &  0 & c  &  d  \\
    \end{pmatrix}, 
\end{align}
where $a, b,c ,d$ are any complex numbers. There are infinitely many solutions. There might be scenarios where some of the Choi matrices $N^T$ corresponding to the solution $N$ are positive, while others are not. We want to examine the compatibility of the data with the pure cause-effect mechanism first, as it is simpler compared to the mixture of the common-cause and cause-effect mechanisms. Therefore, our goal is to find the solutions where $N^{T}$ is the least negative. We thus design the following semidefinite programming to achieve our goal.

Suppose that the solution $N^{T_1}$ has the decomposition $ N^{T}= N^{T}_+ - N^{T}_- $ where $N^{T}_+ \geq 0$ and $N^{T}_- \geq 0$.  The semidefinite programming problem is given by
\begin{eqnarray}
\begin{aligned}
        \text{minimize } &  \Tr N_-^{T} ,\\
    \text{subject to } &\begin{cases} 
    N \, \text{is Hermitian}, \\
    N^{T}= N^{T}_+ - N^{T}_- \\
    \Tr_2 N = \id,\\
    N^{T}_+ \geq 0, N^{T}_- \geq 0.\\
    \end{cases}
\end{aligned}
\end{eqnarray}

Finally,  detailed justifications for protocol for quantum causal inference are listed in the following.

 \begin{figure*}
  \centering
  \includegraphics[width=0.8\textwidth]{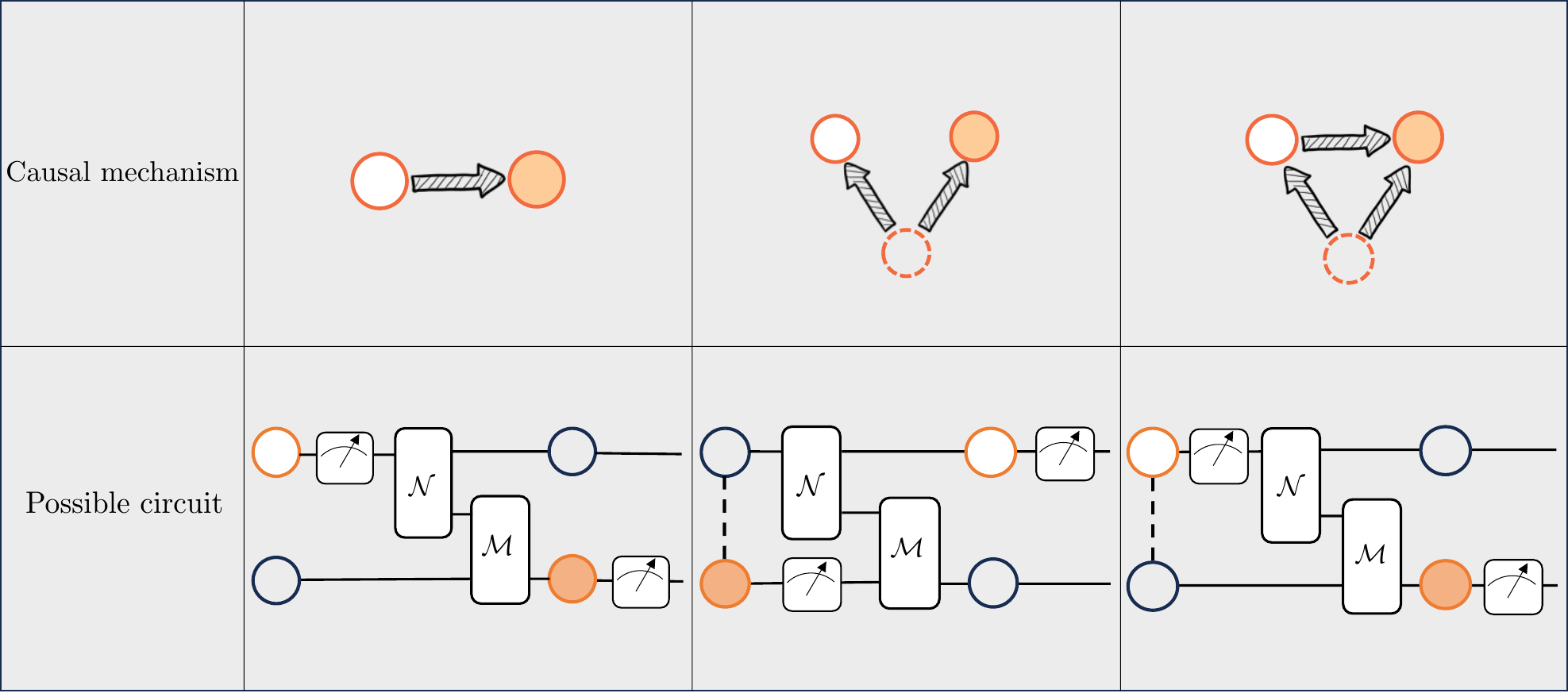}
  \caption{
  { \bf{Correspondence between causal mechanisms and possible circuit models.} }
Each column represents the causal mechanism (top row) and an example of a circuit model realising that mechanism (bottom row). From left to right: (i) the cause-effect mechanism which corresponds to case 1 and in inverse direction case 2 of FIG.~\ref{fig:qsetup}; (ii) The pure common-cause mechanism which corresponds to case 3 in FIG.~\ref{fig:qsetup}; (iii) The mixture of the cause-effect and common-cause mechanisms, which corresponds to case 5 and in inverse direction case 4 of FIG.~\ref{fig:qsetup}. The dashed line in the circuit denotes the initial correlation constituting the common case. The measurements used to define the PDM of interest are also depicted.
}  
  \label{fig:corr}
\end{figure*}

 \begin{itemize}
\item[(1)]
 {\bf Evaluating compatibility with a common-cause mechanism}. Consider the case of no negativity ($f(R_{A_iB_j})=0$). Theorem~\ref{thm:nullneg} shows that negativity associated with two times ($f(R_{A_iB_j})\neq 0$) cannot come solely from a common cause ($R_{A_1B_1}\neq R_{A_1}\otimes R_{B_1}$). Thus in this case the data $R_{A_iB_j}$ is compatible with the (purely) common cause mechanism (case 3 in FIG.~\ref{fig:qsetup}). In particular, by proposition~\ref{eq:RisaRho}, $R_{A_iB_j}=\rho$ where $\rho$ is a single-time standard density matrix so the data generating $R_{A_iB_j}$ can be produced by measurements on two subsystems initially prepared in $\rho$.

\item[(2)] 
 {\bf Evaluating compatibility with different cause-effect mechanisms.}  Consider the case of negativity ($f(R_{A_iB_j})>0$). Theorem~\ref{thm:nullneg} rules out the common cause mechanism and we are left to evaluate the compatibility of the data with cases 1, 2, 4, and 5 in FIG.~\ref{fig:qsetup}. We make use of the time asymmetry results described around Eq.~\eqref{eq:reversePDM} for this evaluation. In particular we extract the two Choi matrices $M^{T}$,  $\bar{M}^{T}$ associated with $R_{A_iB_j}$ and its time reversal $\bar{R}_{A_iB_j}$. The basic idea is that $M^{T}>0$ means there is a CP map on $A$ that gives $B$, indicating that $A$ is the cause and $B$ the effect. More specifically, 
 \begin{itemize}
 \item If $M^{T} \geq 0$ and $\bar{M}^{T} \ngeqslant 0$, the data is compatible with $A \rightarrow B$ (case 1 in FIG.~\ref{fig:qsetup}). In particular,  case 1 in FIG.~\ref{fig:qsetup} corresponds to a semi-causal channel with signalling from $A$ to $B$ acting on a state with no initial correlations. As shown around Eq.~\eqref{eq:forwardCP} the forwards evolution from $A$ to $B$ is then a CP map, such that  $M^{T} \geq 0$. To generate the data $R_{A_iB_j}$ through mechanism 1, one prepares $R_{A_i}$, and applies the CP map associated with $M^{T}$. 
 \item If $M^{T} \ngeqslant 0$ and $\bar{M}^{T} \geq 0$, the data is compatible with $A \leftarrow B$ (case 2 in FIG.~\ref{fig:qsetup}). To generate the data $R_{A_iB_j}$ through mechanism 2, one prepares $R_{B_j}$, and applies the CP map associated with $\bar{M}^{T}$. 
 \item If $M^{T} \geq 0$ and $\bar{M}^{T} \geq 0$, the data is compatible with case 1 and/or case 2 in FIG.~\ref{fig:qsetup}. Now the data can by inspection be generated by either of the immediately above methods. 
 \end{itemize}
 
\item[(3)] If none of the above conditions are satisfied, i.e.,  $f(R_{A_iB_j})> 0, M^{T} \ngeqslant 0$ and $\bar{M}^{T} \ngeqslant 0$, the causal structure is compatible only with case 4 or 5 in FIG.~\ref{fig:qsetup}.  To generate the data using case 4 or 5 one may use a circuit corresponding to that case (circuit (3) in FIG.\ref{fig:corr}), generate the full PDM and then trace out the undesired parts. We do not give a direct simple way to find a circuit consistent with $R_{A_iB_j}$ but there must be one or more such circuits generating the data since we assume the data is generated by either case 1, 2, 3, 4 or 5 and cases 1, 2 and 3 are inconsistent with this data. More specifically, case 3 is ruled out by $f(R_{AB})> 0$, and the cases 1 and 2 are incompatible with $M^{T} < 0$ and $\bar{M}^{T} < 0$ both holding, since the arguments around Eq.~\eqref{eq:forwardCP} show there is a CP map in at least one direction for those cases. 
 
 \end{itemize}

\section{Example: cause-effect mechanism with a common cause} Suppose the PDM corresponds to a semi-causal channel as in FIG.~\ref{fig:semicausal} but with an initially correlated state. One can see that 
\begin{align}
R_{A_1B_2}= 
\frac{1}{2}\begin{pmatrix}
      1& 0 & 0  &  c^2 \\  
      0  &  0 & s^2 & 0 \\
       0 &  s^2 & 0 & 0  \\
      c^2 &  0 & 0 & 1
\end{pmatrix},
\end{align}
is the consequence of the initial state 
$\ket{\psi}_{A_1B_1}= \frac{1}{ \sqrt{2} }( \ket{00}+  \ket{11} ),$
sequentially undergoing swap channel $ \mathcal{N}_{AC} =  S$ followed by $\mathcal{M}_{BC} =e^{i \theta S} := c \id + i s S $, where $C$ is the intermediate system in the semi-causal channel (see FIG.~\ref{fig:semicausal}). (We exclude $s=1$ in which case $A $ replaces $B$ but they do not interact with each other, so the initial correlation does not play a role.)
 
Following our causal inference protocol, we first determine that $f(R_{A_1B_2}) >0,$ implying the existence of the cause-effect mechanism. We then proceed to calculate the Choi matrix of the process and its time reversal,  which can be shown to be 
\begin{align}
M^{T} =
\begin{pmatrix}
  1& 0 & 0  &  s^2 \\  
      0  &  0 & c^2 & 0 \\
       0 &  c^2 & 0 & 0  \\
      s^2 &  0 & 0 & 1
\end{pmatrix}=
\bar{M}^{T} .
\end{align}
Clearly, $M^{T}, \bar{M}^{T} \ngeqslant 0 $.  Therefore the protocol would judge the data to be compatible with case 4 or 5 of FIG.~\ref{fig:qsetup}, corresponding, as desired, to initial correlations and causal influence combined.

\vspace{1cm}
References:
\begin{itemize}

\item[(12)] Fitzsimons, J., Jones, J. and Vedral, V. Quantum correlations which imply causation. Scientific reports 5, 18281 (2016).

\item[(49)] X. Liu, Q. Chen, and O. Dahlsten, Inferring the arrow of time in quantum spatiotemporal correlations, Physical Review A 109, 032219  (2024).

\item[(56)] A.Souza, I.Oliveira, and R.Sarthour, A scattering quantum circuit for measuring Bell’s time inequality: a nuclear magnetic resonance demonstration using maximally mixed states, New Journal of Physics 13, 053023 (2011).

\item[(57)] B. Kraus and J. I. Cirac, Optimal creation of entanglement using a two-qubit gate, Physical Review A 63, 062309 (2001).
\end{itemize}

\end{document}